\documentclass[pdflatex,sn-mathphys]{sn-jnl_nohead}

\usepackage[utf8]{inputenc}
\usepackage[english]{babel}

\title[~]{\vspace{-73 pt} Biaxial strain effects in 2D diamond formation from graphene stacks}

\author*[1,2]{\fnm{Rajaji} \sur{Vincent}}\email{rajaji.vincent@sorbonne-universite.fr}

\author[1]{\fnm{Riccardo } \sur{Galafassi}}

\author[2]{\fnm{Mohammad} \sur{Hellani}}

\author[1]{\fnm{Alexis} \sur{Forestier}}

\author[2,5]{\fnm{Flavio } \sur{Siro Brigiano}}

\author[1,3]{\fnm{Bruno} \sur{Sousa Araújo}}

\author[1]{\fnm{Agnès} \sur{Piednoir}}

\author[1]{\fnm{Hatem} \sur{Diaf}}

\author[2]{\fnm{Fabio} \sur{Pietrucci}}

\author[3]{\fnm{Antonio} \sur{Gomes Souza Filho}}

\author[1,4]{\fnm{Natalia} \sur{del Fatti}}
 
\author[1]{\fnm{Fabien} \sur{Vialla}}

\author*[1]{\fnm{Alfonso} \sur{San-Miguel}}\email{alfonso.san-miguel@univ-lyon1.fr}

\affil[1]{\orgdiv{Université Claude Bernard Lyon 1, CNRS}, \orgname{Institut Lumière Matière ILM UMR5306}, \orgaddress{\city{Villeurbanne}, \postcode{69100}, \country{France}}}

\affil[2]{\orgdiv{Institut de Minéralogie, de Physique des Matériaux et de Cosmochimie}, \orgname{UMR 7590, CNRS, Sorbonne Université, Muséum National d’Histoire Naturelle}, \orgaddress{\city{Paris}, \postcode{75005}, \country{France}}}

\affil[3]{\orgdiv{Departamento de Física, Centro de Ciencias}, \orgname{Universidade Federal Do Ceara}, \orgaddress{\street{} \city{Fortaleza}, \postcode{CEP 60.455-970}, \state{CE}, \country{Brazil}}}

\affil[4]{\orgdiv{Institut Universitaire de France (IUF)}, \country{France}}

\affil[5]{\orgdiv{Laboratoire de Chimie Theorique}, \orgname{Sorbonne Universite}, \orgname{UMR 7616, CNRS, Sorbonne Université}, \orgaddress{\city{Paris}, \postcode{75005}, \country{France}}}

\begin{document}

\abstract{

Discovering innovative methods to understand phase transitions, modify phase diagrams, and uncover novel synthesis routes poses significant and far-reaching challenges. In this study, we demonstrate the formation of nanodiamond-like sp$^3$ carbon from few-layer graphene (FLG) stacks at room temperature and relatively low transition pressure ($\sim$ 7.0 GPa) due to chemical interaction with water and physical biaxial strain induced by substrate compression. By employing resonance Raman and optical absorption spectroscopies at high-pressure on FLG systems, utilizing van der Waals heterostructures (hBN/FLG) on different substrates (SiO$_2$/Si and diamond), we originally unveiled the key role of biaxial strain. 
Ab initio molecular dynamics simulations corroborates the pivotal role of both water and biaxial strain in locally stabilizing sp$^3$ carbon structures at the graphene-ice interface. This breakthrough directly enhances nanodiamond technology but also establishes biaxial strain engineering as a promising tool to explore novel phases of 2D nanomaterials.}

\maketitle

\bmhead{Introduction}

High-pressure has long been utilized as an effective method for inducing phase transformations and discovering novel materials with original properties and technological interest \cite{mao2018solids, mcmillan2002new,miao2020chemistry}. 
Pressure is a directional-dependent thermodynamic quantity in solids, giving rise to isotropic (hydrostatic), biaxial, and uniaxial strain depending on the dimensionality.
The utilization of these physical means, referred to as strain engineering, has recently emerged as a revolutionizing field in material science. 
For instance, uniaxial strain implementation has led to a better understanding and control of catalyst mechanisms \cite{khorshidi2018strain} and exotic electronic transitions \cite{noad2023giant}. 

In the broad field of carbon allotropes, several recent investigations point out to the existence of sp$^2$-to-sp$^3$ transitions in FLG stacks, leading to the formation of two-dimensional (2D) diamond-like materials\cite{lavini2022two, gao2018ultrahard, bakharev2020chemically}. Among them, we focus here on diamondene which is a new 2D form of pure sp$^3$ or mixture of sp$^3$/sp$^2$ phases synthesized through functionalization with OH groups from water under high-pressure conditions \cite{lavini2022two, martins2017raman}. With a unique structure and properties, 2D diamond-like materials offer many potential technological applications such as micro and nano electromechanical systems\cite{krauss2001ultrananocrystalline}, substrates for DNA engineering\cite{yang2002dna}, ultrathin protective coating, electrodes for electrochemical technologies\cite{hupert2003conductive}, spintronics\cite{barboza2011}, biosensors\cite{hartl2004protein,carlisle2004precious}, active laser medium in nano optics, etc. Recently, the use of water as a pressure transmitting medium (PTM) has led to diamondene formation from FLG down to bilayers at a lower transition pressure of $\sim$7 GPa~\cite{martins2017raman,martins2021hard} as compared to the $\sim$15 GPa for the cold compression sp$^2$-sp$^3$ in graphite~\cite{hanfland1989graphite, amsler2012crystal}.
In these seminal diamondene studies, water is considered to play a chemical role in the covalent functionalization of the FLG stacks~\cite{barboza2011, martins2017raman}. 
However, the fundamental transition pathway to diamondene formation
still remains elusive, yet crucial for the advancements of diamond-based nanotechnology. 

High-pressure investigations remain challenging when a 2D nanomaterial is addressed. Issues can originate from sample-to-sample variability, enhanced interface effects, and an anisotropic environment with inevitable physical and chemical interactions between the sample, the (non-)hydrostatic PTM, and the supporting substrate~\cite{bousige2017biaxial,nicolle2011pressure}. This complexity can strongly affect the experimental reproducibility and hinder clear conclusions, especially for atomically-thin 2D materials where strain transfer from the substrate must be taken into account~\cite{machon2018raman,bousige2017biaxial}. To circumvent these challenges, we present an innovative design that compares various local configurations within a single high-pressure run, accompanied by meticulous \textit{in situ} investigation of extrinsic environmental factors.

Though challenging in high-pressure experiments~\cite{pimenta2023pressure,Sun_2021}, we implemented vdWH, namely FLG covered with hexagonal Boron Nitride (hBN), to locally modulate the interface interaction with PTM water and demonstrate the direct chemical role of water.
With high-pressure runs performed on SiO$_2$/Si and diamond substrates, we reveal the need for strong in-plane biaxial strain (physical role) induced by a substrate to form diamondene at room temperature and relatively low pressure.
\textit{In situ} optical characterization of the pressure-induced phase is supported by {\sl ab initio} molecular dynamics (MD) simulations, whose results advocate for sp$^2$-to-sp$^3$ bond formation initiated on nucleation sites and promoted by both H$_2$O local functionalization and biaxial strain. With the use of these newly identified synergistic environmental means, our work offers new pathways for the synthesis of nanomaterials at originally low pressure by exploiting strain-mediated chemistry and phase transition.

\bmhead{Sample design and fabrication}

\begin{figure}[h]%
\centering
\includegraphics[width=1.0\textwidth]{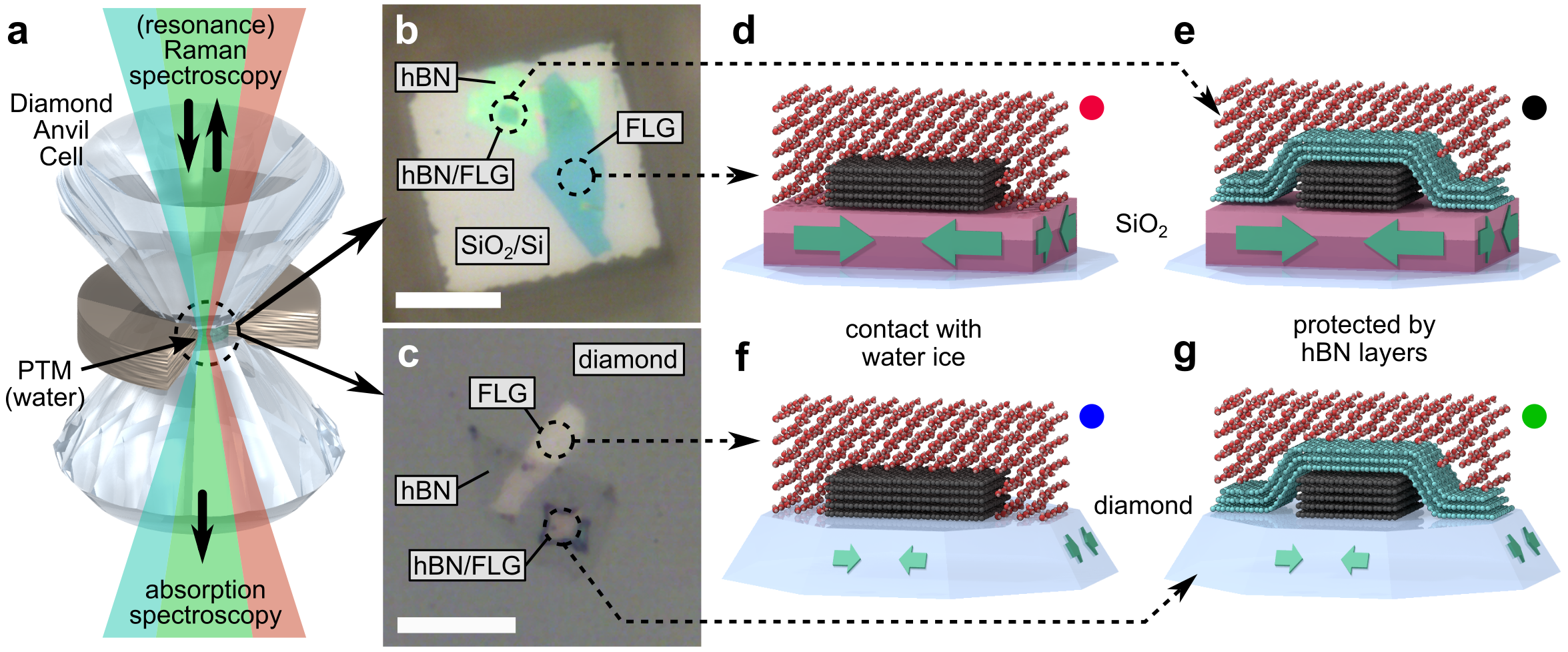}
\caption{(a) Schematic representation of the DAC probed with optical spectroscopies. (b,c) Optical images of the prepared FLG, hBN, and vdWH (hBN/FLG) samples transferred onto (b) SiO$_2$/Si and (c) diamond substrates. Scale bars are 30 $\mu$m. (d-g). Schematic representations of the FLG sample (black atoms) on the four explored configurations with both unprotected and hBN (blue atoms) fully covered regions in contact with the solid ice VII (red-white molecules), and both SiO$_2$/Si (pink) and diamond (light blue) substrates. The green arrows indicate different biaxial compression resulting from substrate deformation in high-pressure conditions. The dots (red, black, blue and green) define the color code used in the following figures.}\label{fig1}
\end{figure}

To unravel the high-pressure phase transition conditions with robust evidence in a diamond anvil cell (DAC, Fig.~\ref{fig1}a), we propose to probe and compare with \textit{in situ} characterization FLG samples in different local configurations under high-pressure. 
We innovatively designed two experimental strategies, taking advantage of (i) the biaxial strain induced by SiO$_2$/Si and diamond substrates supporting the FLG, and (ii) the interface passivation introduced by hBN van der Waals capping of the FLG. By combining these approaches, we explored four local configurations in two high-pressure runs (Fig.~\ref{fig1}b-g), which allowed us to disentangle the physical and chemical effects involved in diamondene formation.

Under high-pressure conditions, supported FLG mechanical evolution depends on the biaxial compression from the substrate rather than directly the pressure transferred by the PTM, and it thus deviates significantly from bulk graphite behaviour~\cite{machon2018raman, bousige2017biaxial}. The effective strain experienced by FLG is related to the intrinsic substrate compressibility and the efficiency of strain transmission at the interface. By using diamond and SiO$_2$ as substrates, given that the bulk modulus of diamond is one order of magnitude larger than for SiO$_2$, we choose two discriminating configurations to explore weak and strong, respectively, biaxial strain induced in the FLG samples~\cite{bousige2017biaxial}.

Stacks of different layered van der Waals materials such as graphene and hBN in so-called vdWH present perfect contaminant-free interfaces with atomic conformation \cite{geim2013van}. Capping or encapsulation of graphene with hBN, a large bandgap dielectric, is a well-recognized approach to passivate its surface and thus preserve its intrinsic properties \cite{wang2013one}. Here, we use such a scheme for our FLG samples to prevent any chemical interaction with water as the PTM, while still ensuring good transmission of the mechanical strain from the high pressure. We fabricated vdWHs made of both unprotected and fully covered FLG deposited using dry transfer either directly on the diamond anvil surface (Fig.~\ref{fig1}c) or on a small SiO$_2$/Si substrate priorly glued to the anvil (Fig.~\ref{fig1}b). 
The Raman spectroscopy and atomic force microscopy (AFM) characterization at ambient conditions of the prepared vdWHs ($\sim$20 nm-thick hBN on 15 to 100 nm-thick FLG) samples on the two different substrates are shown in Supplemental Materials (SM, see Fig.S1-5). 

\bmhead{Raman spectroscopy under high-pressure}

\begin{figure}[t]%
\centering
\includegraphics[width=1\textwidth]{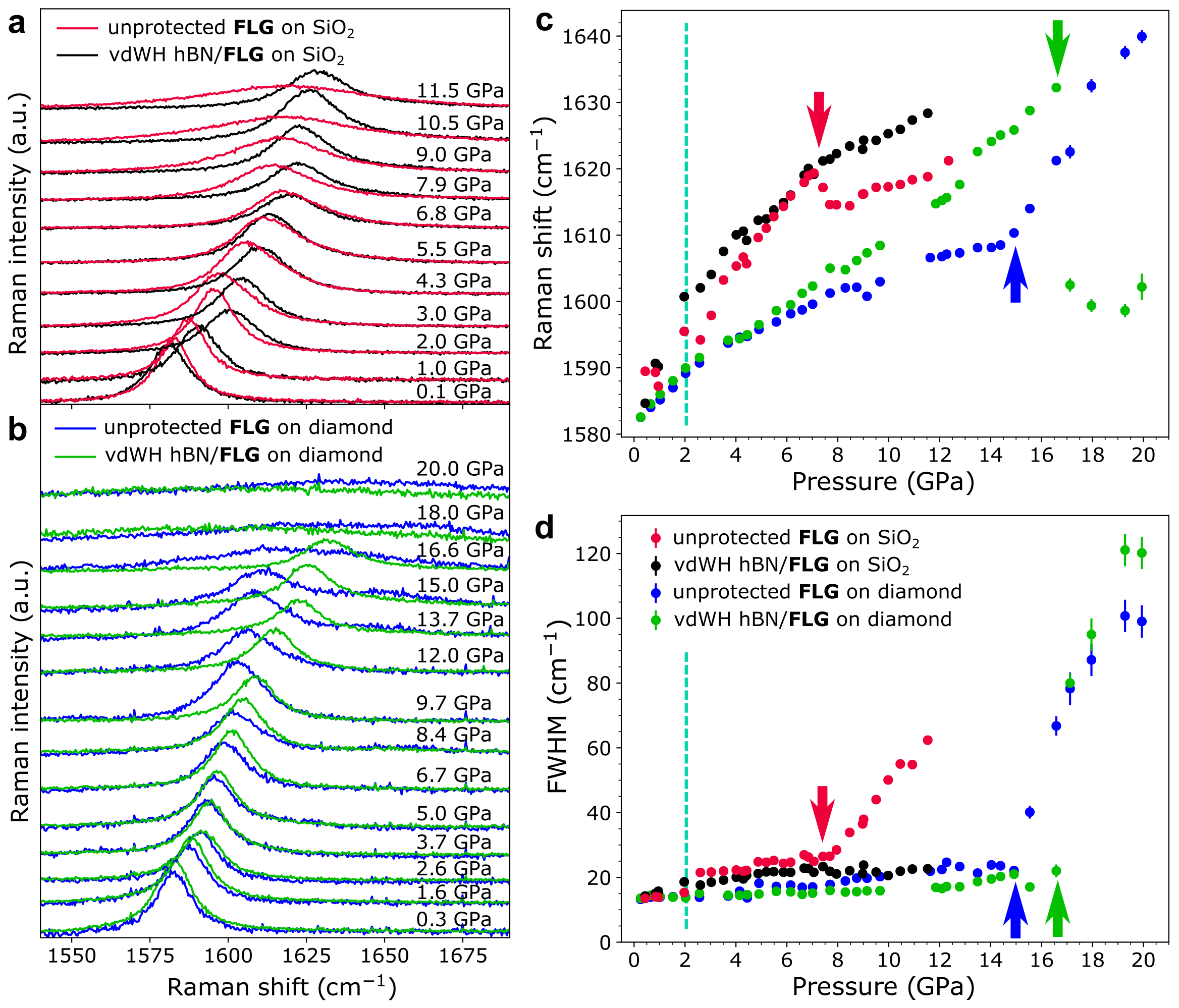}
\caption{
Pressure evolution of the G band Raman spectra of FLG samples for the unprotected and fully covered with hBN regions deposited on (a) SiO$_2$/Si and (b) diamond substrates. Pressure dependence of the G band's (c) Raman shift and (d) FWHM is extracted for all four configurations presented in Fig.~\ref{fig1}. The dotted vertical line marks the environmental transition from water to ice VII while arrows indicate the intrinsic transitions for FLG to a sp$^3$ phase at the lowered (7 GPa) and usual bulk graphite ($\gtrsim$ 15 GPa) pressure thresholds. All data are obtained with a 532 nm laser wavelength excitation, and utilizing pure water as the PTM.}\label{fig2}
\end{figure}

During the high-pressure runs in the DAC, we performed \textit{in situ} Raman spectroscopy (532 nm wavelength) on FLG in all four configurations of capping and substrate (see Fig.~\ref{fig1}d-g). We focused our study on the G band, a first order Raman active mode (E$_{2g}$ symmetry) specific to sp$^2$ bonds in graphite and graphene structures\cite{ferrari2013raman}. Other relevant modes from the FLG are fully or partially masked by the diamond anvils signal and shown in SM (Fig.S6,7).
The evolution of the G band with pressure is presented in Fig.~\ref{fig2}a,b with a quantitative analysis of the Raman shift (Fig.~\ref{fig2}c), and full width at half maximum (FWHM) [Fig.~\ref{fig2}d] using Lorentzian fitting. 
Additional spectra, and their extracted data are presented in SM (Fig.S8-12).  

In all configurations upon increasing pressure, a similar trend is observed on the G mode with first a monotonous hardening (i.e. blue shift) of the Raman shift, and an almost constant or slightly increasing linewidth. This behaviour corresponds to simple transmission of strain with no structural transformation in the FLG samples. 
At higher pressure, a clear change in behaviour is evidenced with a progressive disappearance of the G band associated with a break in the Raman shift slope and a strong broadening of the linewidth (arrows in the Fig.~\ref{fig2}c,d). As will be discussed in the following, this transition can be attributed to a structural change in the FLG from sp$^2$ to sp$^3$ bond structure. 
At pressures $\gtrsim$ 15 GPa for the samples on diamond substrate, both unprotected and covered FLG exhibit the expected transition of graphite from hexagonal to monoclinic phase (m-carbon) under cold compression ~\cite{hanfland1989graphite, amsler2012crystal}.
Importantly, we observe that the threshold pressure for a similar transition is significantly lowered down to $\sim$7 GPa only in the configuration of unprotected FLG supported on SiO$_2$/Si. 
In contrast, the case of hBN-covered FLG supported on SiO$_2$/Si shows no transition in the studied range up to 12 GPa, which corresponds to the structural loss of the substrate (see below and SM, Fig.S14).
The direct comparison of all four configurations clearly demonstrates the combined effect of water ice VII at the top interface and of SiO$_2$/Si supporting substrate at the bottom one. 

\begin{figure}[h]%
\centering
\includegraphics[width=1\textwidth]{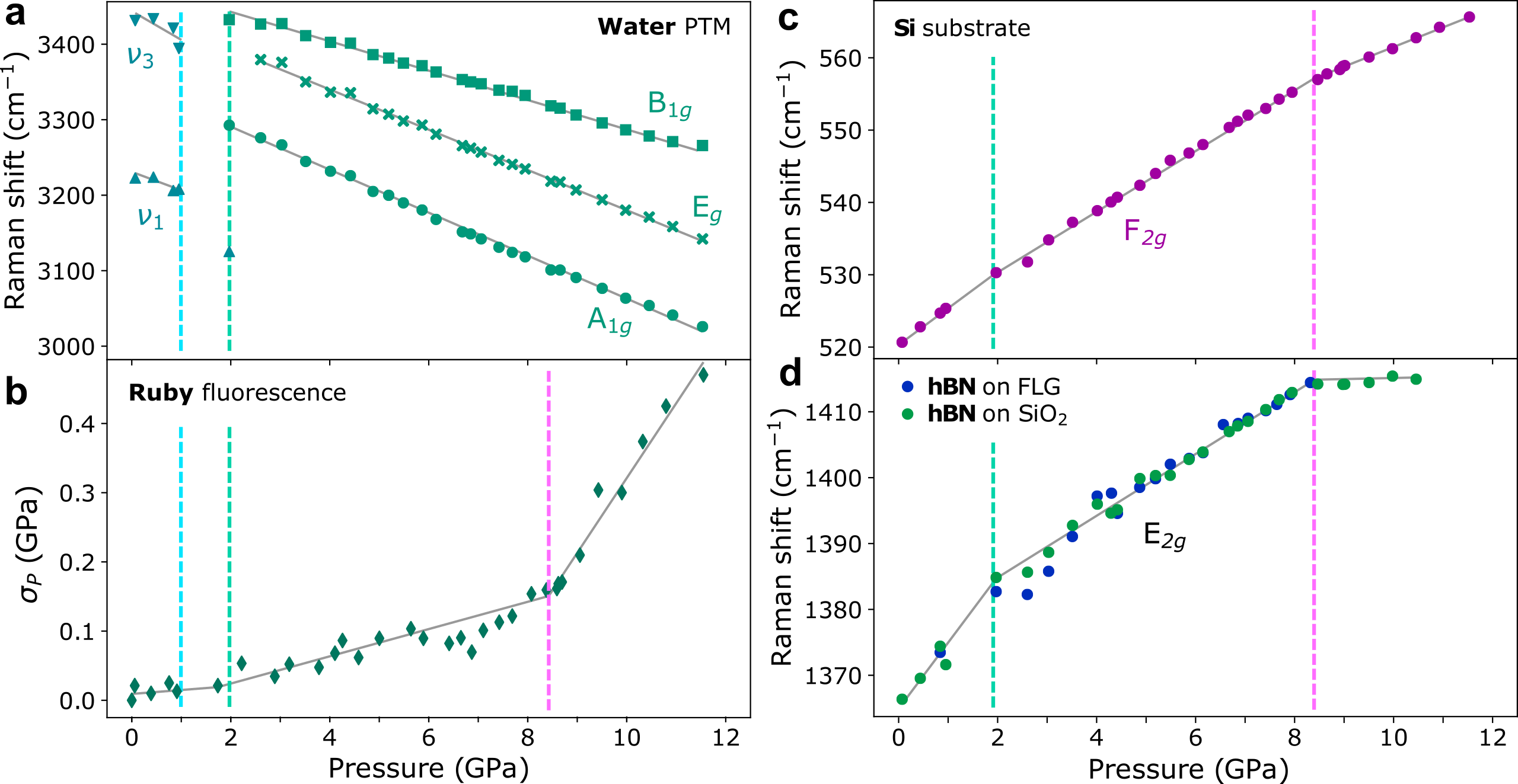}
\caption{\textit{In situ} high-pressure investigations of environmental effects during the SiO$_2$/Si substrate experiment.
Pressure dependence of (a) the various Raman modes' shift of water PTM, (b) the standard deviation $\sigma_P$ of the pressure inside the chamber extracted from the fluorescence of four different rubies, (c) the Raman shift of the Si substrate $F_{2g}$ mode, and (d) the Raman shift of the hBN flake $E_{2g}$ mode in regions supported on the SiO$_2$/Si substrate and on top of the FLG in the vdWH. Symmetry denomination of the respective modes are indicated. The solid lines are guides to the eye. Vertical dotted lines mark the pressures where changes in behaviours are observed. We emphasize here that no environmental effect is correlated with the observed FLG transition at the lowered pressure of $\sim$ 7 GPa.}\label{fig3}
\end{figure}

Other minor changes in the Raman G band features as pressure increases can be attributed to extrinsic environmental effects.
Water undergoes a pressure-induced structural phase transition from liquid to a solid phase [Ice VI, space group(SG): P4$_2$/nmc)] at $\sim$1.0 GPa and then another solid phase [Ice VII, SG: Pn3m] at $\sim$2.0 GPa
\cite{pruzan1990raman, zha2016new}. This behaviour is directly confirmed in our study through Raman spectroscopy of the PTM water depicted in Fig.~\ref{fig3}a and SM (Fig.S15). The discontinuity observed in FLG signals at 2.0 GPa in Fig.~\ref{fig2}c,d is attributed to this environmental PTM phase solid-to-solid transition~\cite{Forestierjpcc2020}. All configurations show such effect, thus validating that FLG is affected by the strain stemming from the PTM, notably even in the hBN-covered cases.  
Non-hydrostaticity is another effect emerging from the solid phase transition of water as PTM. This is a source of disorder which can induce the slight but steady increase of the FLG G band linewidth above 2 GPa (Fig.~\ref{fig2}d). It can also explain the slightly different transition pressures observed for the FLG samples on diamond (Fig.~\ref{fig2}c).
This non-hydrostatic situation might also lead to more local structural modification such as shearing or rippling of graphene layers. To address this question, we carefully loaded up to four micron-scale ruby crystals in the high-pressure chamber in our runs and monitored \textit{in situ} their fluorescence as pressure markers (SM, Fig.S17). We used the standard deviation of the locally extracted pressures as a sensitive estimate of the global non-hydrostaticity~\cite{klotz2009hydrostatic}. In Fig.~\ref{fig3}b, we observe a steady yet reduced rise of non-hydrostaticity after the solidification, and identify the onset of a strong increase only above 8 GPa. Importantly, we do not find a behaviour correlated with the anomaly in FLG observed at $\sim$ 7 GPa, thus dismissing this extrinsic influence in this transition. 

In the pressure range of 2-7 GPa, as evidenced from Fig.~\ref{fig2}c, the FLG G-band frequency increases with pressure at a higher rate with the SiO$_2$/Si substrate when compared to diamond substrate, for both unprotected and covered FLG signals. This result confirms the intended dominant role of biaxial compression from the two supporting substrates, as designed.
Si presents a first order structural phase transition from diamond cubic (Si-I, SG: Fd-3m) to $\beta$-Sn structure (Si-II, SG: I4$_1$/amd) expected at $\sim$12 GPa \cite{zeng2014phase}. This is directly confirmed with a strong modification in morphology of the SiO$_2$/Si substrate observed in optical microscopy (SM, Fig.S14). This transition prevented further pressure increase in the case of SiO$_2$/Si supported FLG samples, and did not allow to reach the expected sp$^2$ to sp$^3$ transition above 15 GPa for the hBN-covered FLG.   
\textit{In situ} Raman spectroscopy of the Si substrate's F$_{2g}$ mode, shown in Fig.~\ref{fig3}c and SM (Fig.S14) confirms the structural transition with a complete loss of the signal at $\sim$ 12 GPa. 
The observed anomaly in the Raman shift evolution with pressure at $\sim$2.0 GPa corresponds to the PTM solid phase (ice VII) transition.
The anomaly at $\sim$8.5 GPa might be directly or indirectly correlated with the observed non-hydrostaticity build-up (Fig.~\ref{fig3}b). 
Further analysis and discussion of these environmental factors can be found in the SM.

Finally, Raman spectroscopy of the E$_{2g}$ mode of hBN, probed in regions supported on SiO$_2$/Si and covering FLG in the vdWH (Fig.~\ref{fig3}d and SM, Fig.S16) presents anomalies only at pressures related to the previously discussed environmental effects. The similar behaviour of hBN supported on SiO$_2$/Si and on FLG, especially around 7 GPa, demonstrates the absence of chemical interaction between hBN and FLG inside the vdWH under high-pressure conditions. Therefore, by ruling out all the possible environmental artifacts stemming from the PTM, the substrate and the hBN-covering, our understanding of the observed features in the Raman G band's high-pressure evolution clearly supports that FLG undergoes an intrinsic structural transition around 7 GPa.
To trigger this transition, we bring evidence for the crucial role of chemical interaction of unprotected FLG with water ice VII when comparing to the hBN-covered configuration, and of physical biaxial strain when comparing the SiO$_2$/Si to the diamond supported configuration.

\bmhead{Characterization of the high-pressure phase}

Characterizing the new phase \textit{in situ} is necessary, given that the discussed Raman features show a reversible behaviour when pressure is released (see SM, Fig.S8-13), hinting that the produced phase is not stable in ambient conditions. However, it is  challenging due to the similar carbon atoms bonding nature of the DAC and the expected diamondene. 
The anticipated sp$^3$ phase signature in Raman spectroscopy typically appears within the 1300-1350 cm$^{-1}$ range and coincides with the intense first-order Raman peak of diamond from DAC (see SM, Fig.S1,4), thus preventing a direct spectroscopic identification. 
Broadening of the G band linewidth is generally considered a signature of a structural phase transition for the case of graphite\cite{amsler2012crystal, wang2012crystal}. Although complete disappearance is expected for a fully sp$^3$ structure, linewidth can first be increased due to electron-phonon coupling in the newly created sp$^3$ bonds. This behaviour is coherent with our observation of FLG transition at 7 GPa (Fig.~\ref{fig2}d). 

\begin{figure}[h]%
\centering
\includegraphics[width=.9\textwidth]{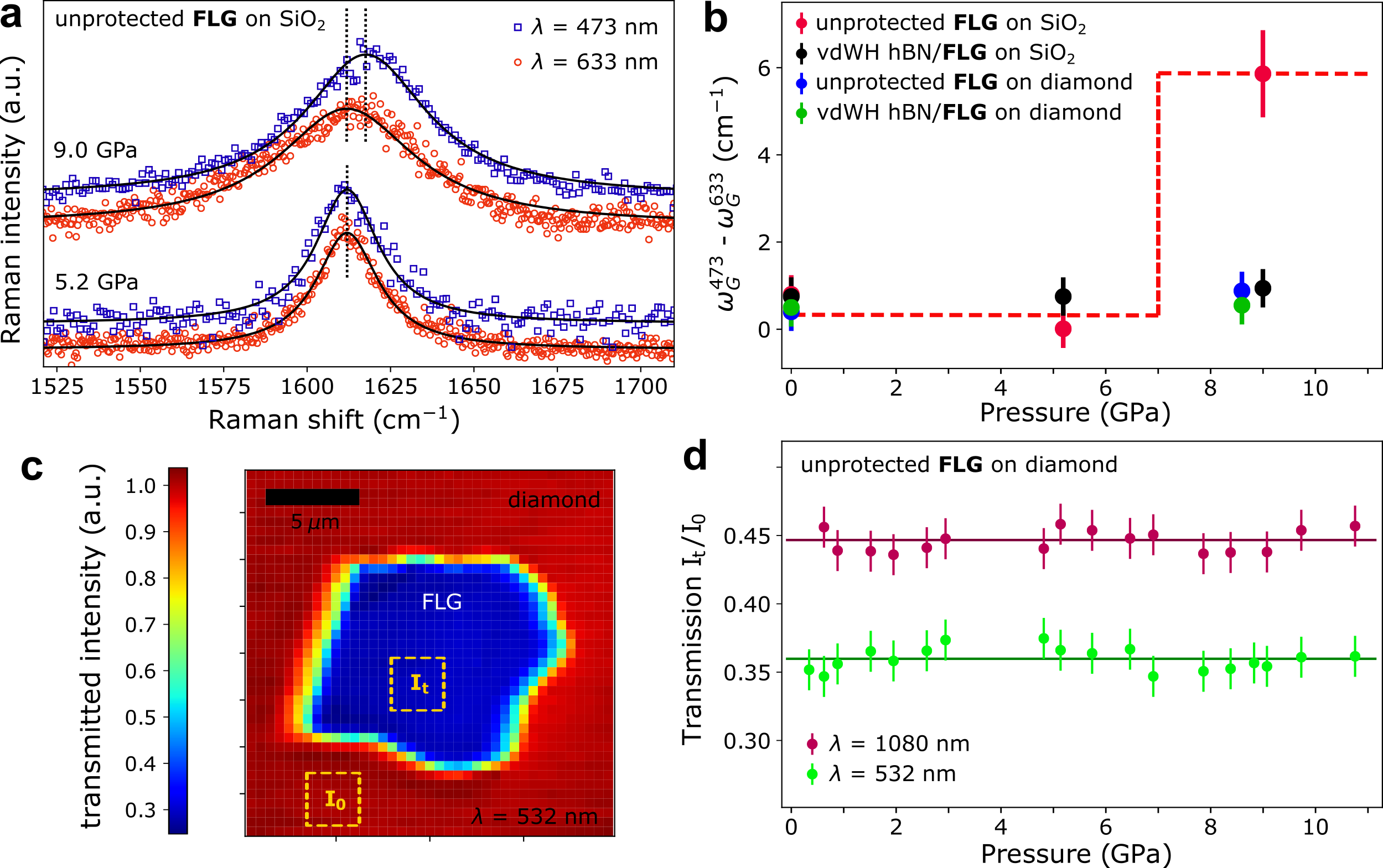}
\caption{(a). Resonance Raman spectroscopy of unprotected FLG samples on the SiO$_2$/Si substrate. Raman spectra of G band are shown before (5.2 GPa) and after (9.0 GPa) the transition pressure value for two laser wavelengths of 473 (blue square) and 633 (red circle) nm. The solid black line corresponds to Lorentzian line shape profile. Dotted vertical lines highlight the dispersive behaviour. (b). Difference of the G band Raman shift measured with 473 and 633 nm laser wavelengths for all four investigated configurations at selected pressures. Only the configuration of FLG in direct contact with water and supported on SiO$_2$/Si substrate shows a dispersive behaviour above 7 GPa, as highlighted with the red dashed guide to the eye. (c) Optical transmission map of a FLG sample on the diamond substrate. The dotted yellow square areas represent the regions of reference for FLG and diamond background signal evaluation, I$_t$ and I$_0$ respectively. Laser wavelength is 532 nm and scale bar is 5 microns. (d) Pressure-dependent variation of relative transmission I$_t$/I$_0$ for two different incident laser wavelengths (532 and 1080 nm). The solid lines are guides to the eye.}\label{fig4.jpg}
\end{figure}

The wavelength dispersive nature of the G band appears as another key experimental indication for a mixed sp$^3$/sp$^2$ phase, i.e. diamondene formation. While G band's Raman shift is independent of the excitation wavelength in graphite \cite{ferrari2001resonant,vidano1981observation}, breakdown in the Kohn anomalies of the phonon dispersion in diamondene phase has been proposed to yield a dispersive behaviour~\cite{martins2017raman}.
We probed the dispersion of G band's Raman signal using incident wavelengths of 473 and 633 nm at selected pressure values. Representative spectra are shown along with the extracted dispersion in Fig. \ref{fig4.jpg}a,b and SM (Fig.S18).
In all configurations of capping and substrate, no dispersion at pressures below the identified structural transition is observed, as expected for the pure sp$^2$ structure in FLG. In both cases with diamond substrate, and with hBN-covered FLG supported on SiO$_2$/Si, this behaviour remains at a pressure of $\sim$ 9 GPa. Only in the FLG supported on SiO$_2$/Si substrate with an unprotected interface in contact with water, the G band shows a clear dispersive behaviour (Fig. \ref{fig4.jpg}b). This correlated observation further confirms that a structural transition occurred around 7 GPa in this configuration only, with here a strong hint towards the formation of bonds with sp$^3$  character.  
We note that similar G band dispersive behaviour induced at lower pressures with water PTM has also been reported for FLG on highly compressible Teflon substrate~\cite{martins2017raman}, confirming the role of transmitted biaxial strain.

Our original observation of prevented transition with water PTM when supported on weakly compressible diamond has been confirmed with complementary \textit{in situ} optical absorption measurements.
Relative transmission through and away from unprotected FLG samples under high-pressure is measured with two different incident wavelengths (532 and 1080 nm) and reported in Fig. \ref{fig4.jpg}c,d and SM (Fig.S19). We find relative transmission values in accordance with the FLG thickness, with an expected slightly higher value at longer wavelength \cite{hanfland1989optical}. We observe a constant behaviour when pressure is increased up to 11 GPa for both wavelengths. By considering the strong influence on optical properties of hybridization characters (sp$^2$ and sp$^3$) in graphite \cite{hanfland1989optical,utsumi1991light}, this further demonstrates the structural stability of the hexagonal phase and the absence of transition around 7 GPa even with exposure to water PTM. Therefore, the requirement of applying strong biaxial strain to induce the emergence of sp$^3$ bonding is here confirmed. However, this specific configuration has not been characterized with optical transmission due to the SiO$_2$/Si substrate’s opaqueness as opposed to transparent diamond.

The fundamental impact of strong biaxial strain in FLG is to reduce the C-C in-plane distance~\cite{bousige2017biaxial}, which could favor a pyramidalization of the orbitals. 
We propose that biaxial strain in FLG helps to locally stabilize sp$^3$ carbon structures initiated by the functionalization of water molecules at the graphene-ice interface, allowing a lower pressure transition. 
This mechanism is addressed in the following on a bilayer model system to reduce computation complexity.

\bmhead{Atomistic modeling}

\begin{figure}
\includegraphics[width=.95\textwidth]{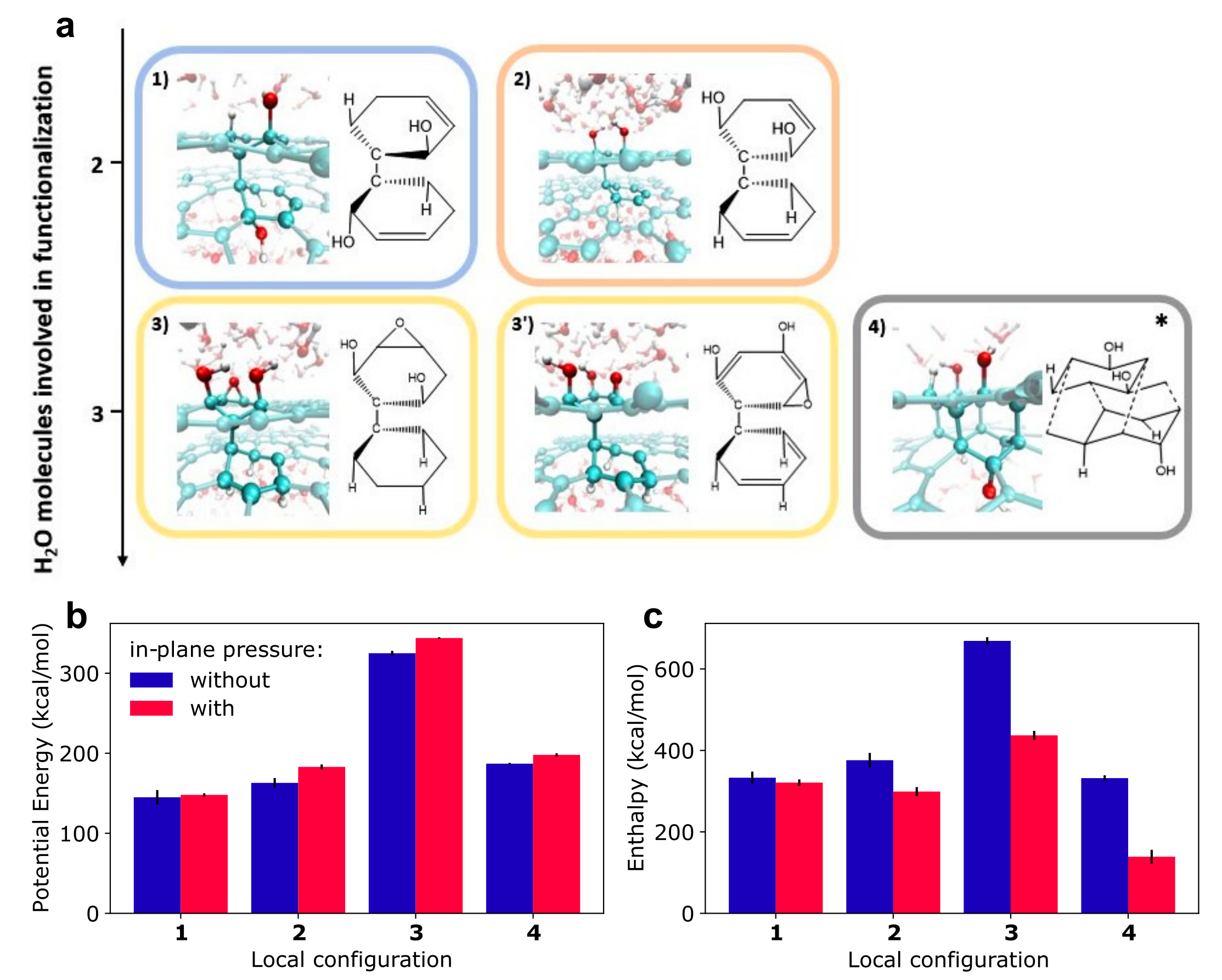}
\caption{(a) Bilayer graphene in contact with ice VII at 300~K: configurations that proved stable after several picoseconds of {\sl ab initio} MD simulation. Configurations are tested stable under 6.8 GPa with and without in-plane pressure of 15 GPa. Structure 3 and 3' are MD simulation result from same structure with and without in-plane pressure, respectively. C-C interlayer bonds in structure 4, marked with a star, are formed spontaneously during geometry optimization and MD simulation.
Corresponding calculated (b) potential energies and (c) enthalpies in all 4 configurations with (red) and without (blue) in-plane pressure applied. The data are reported in Table S1 of the SM.}
\label{fig5}
\end{figure}

We performed {\sl ab initio} Born-Oppenheimer molecular dynamics (BOMD) simulations of bilayer graphene (BLG) at the interface of a ice VII matrix under high pressure (6.8 GPa, 300K), with and without biaxial strain (introduced here with a higher in-plane pressure of 15 GPa). 
We investigated the interplay between BLG functionalization at the interface with ice, due to reactive H$^+$ and OH$^-$ ions from water dissociation, and the formation of local sp$^3$ diamond-like structures with interlayer C--C bonds. 
Following experimental spectroscopic evidence of short-range domains (nano-patches) characterized by proton ordering in the proton-disordered water ice VII phase \cite{walrafen1982raman,pruzan1990raman}, we modeled ice VII/BLG interface with partial proton ordering in the first layer (see Computational Methods). 
Such ice VII proton-ordered structures were systematically found stable when simulated by BOMD at 6.8 GPa.
We generated several different initial structures of functionalized BLG, with CH, COH and/or COC groups~\cite{dreyer14,mouhat20,guo2022controlling} formally corresponding to splitting of two and three water molecules, as shown in Fig.~\ref{fig5}a.
Carbon interlayer bonds were either introduced by us in the initial model or formed spontaneously during geometry optimization and BOMD simulation. 
All structures displayed sp$^3$ carbon arrangements that proved stable within several picoseconds of BOMD simulation, with and without in-plane pressure.
Note that, on the contrary, in absence of functional groups, the C--C interlayer bonds are unstable and spontaneously break within less than 1 ps.

The energetics of the different ice VII/BLG interface configurations are reported in Figure~\ref{fig5}b,c and SM (Table S1). They provide original insights about the effect of type and position of functional groups on the stability of C--C interlayer bonds. 
In particular, they confirm that in-plane pressure, with an effect similar to biaxial strain induced by a substrate in our experiments, leads to a significant decrease in enthalpy in all cases.
From the energetic viewpoint, it is also important to underline that, under the in-plane pressure conditions, passing from one C--C interlayer bond (structures 1, 2, 3) to 4 bonds (structure 4) leads to a strong enthalpy decrease. 
These configurations might thus represent the early stages of a nucleation process leading from small-range modifications of the sp$^2$ graphene lattice to the growth of a long-range sp$^3$ diamondene structure in FLG.

\bmhead{Conclusion}

In summary, we performed technically challenging high-pressure experiments with \textit{in situ} optical spectroscopies on FLG samples tuning their interfaces: interacting directly with water or passivated with hBN, and with weak or strong strain induced by diamond or SiO$_2$/Si substrates, respectively.
Clear evidence for intrinsic sp$^2$-to-sp$^3$ structural transition at a pressure of 7 GPa, significantly lower as compared to bulk graphite, has been demonstrated only when graphene layers are in direct contact with water PTM and SiO$_2$/Si substrate.
The present experimental work corroborated with {\sl ab initio} MD simulations demonstrates the intricate interplay between (physical) biaxial strain and (chemical) water interaction in driving the FLG to transform into diamondene under high-pressure conditions.
In the pursuit of both stabilizing and lowering the transition pressure for the diamond-like carbon phase, our findings introduce a new tuning parameter which is in-plane biaxial strain. 
Our results not only expand the frontiers of diamondene synthesis but also pave the way for extending this biaxial strain concept to engineering of distinct phases in other 2D and nanomaterials, and potentially reshaping the landscape of pressure-induced phase transitions in supported atomically-thin layered materials.

\section*{Methods}\label{sec11}

\bmhead{Sample preparation}
The FLG and hBN samples were prepared by the standard mechanical exfoliation method from Highly Ordered Pyrolytic Graphite (HOPG) and high-quality grown single crystals of hBN~\cite{Maestre_2022}, respectively. Following the scotch tape exfoliation process, the resultant samples were initially transferred to a temporary substrate, such as Polydimethylsiloxane (PDMS). After that, the required FLG and hBN samples were carefully identified by optical contrast of the microscope. We have used SiO$_2$/Si substrates consisting of a 300 nm thick layer of amorphous oxidation onto 50 $\mu$m thick silicon wafers. Using diamond saw, the large SiO$_2$/Si substrate was cut into small pieces of cube shapes (60 $\mu$m x 60 $\mu$m x 50 $\mu$m). This small-size SiO$_2$/Si substrate will be placed in the center of the diamond culets surface and glued. Subsequently, the suitable FLG sample is transferred from the PDMS substrate to a small SiO$_2$/Si substrate using a customized setup which works on the principle of the deterministic dry transfer method~\cite{castellanos2014deterministic}. Following this, a precisely sized thin layer of hBN is carefully positioned onto the upper surface of the FLG sample, culminating in the creation of the desired vdWHs (hBN/FLG) sample. In the case of the diamond-based experiments, the transfer of FLG and hBN samples was done directly on one of the diamond culet surface [oriented as (1 0 0)] of the anvils within the DAC setup. 

\bmhead{Atomic Force Microscopy (AFM)}
AFM characterization was done using a commercial AFM setup (MFP-3D, Asylum Research, Oxford Instruments). The height profile analysis reveals the thickness of the prepared samples and the details are commented in the SM.

\bmhead{Raman scattering measurements}
In this study, we have used two different commercial Raman spectrometers.  (i). LabRAM HR evolution 800 with the wavelength of $\lambda$ = 532 nm (Nd-YAG solid-state laser) (ii). Horiba Aramis spectrometer with the wavelength of $\lambda$= 473 nm (DPSS laser) and 633 nm (He Ne laser). Both spectrometers have a spectral resolution of about 1.0 cm$^-$$^1$ and operate in backscattering configurations. For the high-pressure experiments, we have used 50X objectives, 150 $\mu$m entrance slit width, a grating of 1800 groves/mm, and the backscattering Raman signal was detected by high-sensitivity CCD detector. The Raman measurements were carefully performed with low laser powers (less than 1.0 mW) to avoid the risk of the sample damage due to laser heating effects. During the high-pressure experiments, the typical data accumulation time for each of the collected Raman spectra is about 5 min.

\bmhead{Optical Absorption measurements}
Absorption spectroscopy was performed using either a 532 nm CW laser diode or a 1080 nm pulsed Ti:Sa source. The DAC was placed in-between a transmission optical system with two objectives (Mitutoyo 100X, 10 mm working distance, corrected for the diamond anvil aberration), and followed by an avalanche photodiode (Hamamatsu). Coupled with a 3-axis piezo-positioner stage (PI), it allows optical transmission mapping of the sample inside the DAC at a diffraction limited resolution.  

\bmhead{High Pressure experiments}
The \textit{in situ} high-pressure Raman scattering and optical absorption experiments were carried out using the membrane-type DAC with an anvil culet of 400 $\mu$m in diameter. The Rhenium and stainless steel (T301) type gasket materials with starting thickness of $\sim$200 $\mu$m were used. It was preindented to $\sim$120 $\mu$m thickness and the hole in diameter of $\sim$150 $\mu$m was drilled at the center of the gasket, which served as the sample chamber. The drilling of the hole was achieved using a BETSA drilling machine which works on the principle of Electric Discharge Machining (EDM). Pure water (H$_2$O) served as the PTM, maintaining a hydrostatic environment up to around 1.0 GPa. Beyond this threshold, quasi-hydrostatic or non-hydrostatic conditions emerged within the pressure chamber due to the formation of ice. Several small size ($\sim$5 $\mu$m) ruby spheres were strategically distributed within the sample chamber to serve as pressure markers. The \textit{in situ} pressure was determined by the standard ruby luminescence method.

\bmhead{Computational methods}
Orthorhombic models have been adopted, composed by BLG  with AB initial stacking geometry, forming an interface with ice VII (001 facet): the latter is proton-disordered and obeys the ice rules with zero net polarization (generated using the GenIce software~\cite{matsumoto2018genice}). 
The models include 120 carbon atoms and 128 water molecules.
Simulations were performed in the PBE approximation of density functional theory~\cite{Perdew1996} (spin-restricted) together with Grimme’s (D3-BJ) correction for dispersion interactions~\cite{Grimme2010,Grimme2011}, GTH pseudopotentials and combined plane-wave (850 Ry cut-off) and DZVP-MOLOPT SR (short range) basis sets~\cite{Goedecker1996}.  
 
The position and cell parameters were optimized imposing static $P_{zz}=6.82$~GPa, 
$P_{xx}=P_{yy}=100$~KPa (ambient pressure).    
The resulting cell has dimensions $12.48\times 12.91\times 19.65~\AA$; further simulations were performed at constant volume.
A snapshot of the resulting simulation box is shown in the SM (Fig.S21).
A second model of ice VII/BLG with partial proton ordering at the interface was generated. Details regarding the model properties and the protocol for its generation  can be found in the SM.
BOMD simulations were performed with a time step of 0.5~fs. We adopted the Nosé-Hoover thermostat~\cite{Nose1984,Nose2006} to control the average temperature 300 K in the $NVT$ ensemble. 
The length of the Nose-Hoover chain was equal to 3, whereas the time constant of the thermostat was set to 100~fs, with  3rd order Yoshida integrator and multiple time step set to 2. 
Trajectories of 5~ps were found to be sufficiently long to reach the target temperature.

\backmatter

\section*{Supplementary information}

The online version
contains supplementary material available at ...

\section*{Data availability}
The data supporting the findings of this study are presented in the main text and the supplementary material. Further data are available from the corresponding authors upon reasonable request.

\section*{Acknowledgments}

The authors sincerely thank Pr Catherine Journet and Camille Maestre (LMI, Lyon) for providing the hBN flakes and Dr. Stefan Klotz (CNRS, IMPMC, Sorbonne University, Paris) for the discussion about the standard deviation of pressure transmitting medium and water-ice phase transitions. The authors acknowledge the financial support from ANR and project 2D-PRESTO ANR-19-CE09-0027. FV acknowledge the financial support from IDEXLYON through the IMPULSION program. We sincerely acknowledge the various help from SOPRANO and FemtoNanoOptics (ILM) team members. The authors acknowledge CAPES-COFECUB agreement for partial support. ACSF acknowledge Brazilian agency CNPF for funding. This work was performed using HPC resources from GENCI (grants 2022-A0130811069 
and  2021-A0110811069).

\section*{Author contributions}
R.V. and A.S.M. designed the study; A.S.M. and F.V. supervised the project, with critical inputs from N.F. and A.G.S.F.; R.V. prepared the samples with the contributions from R.G. and F.V.; A.F. contributed to the development of the experimental techniques; R.V. performed the high pressure spectroscopy measurements with the contributions from R.G., F.V., H.D. and B.S.A; R.V., R.G. and F.V. analyzed the experimental data; A.P. performed the AFM measurements; M.H., F.S.B. and F.P. performed the MD calculations; F.P. supervised the theoretical part of the project; F.V., R.V. and A.S.M. wrote the manuscript, with critical revision from all authors. The manuscript reflects the contributions of all authors.

\section*{Competing interests}
The authors declare no competing interests.

\bibliography{biaxial.bib}

\end{document}